\documentclass[preprint2,10pt]{aastex} 
\def\ls{\mathrel{\raise0.27ex\hbox{$<$}\kern-0.70em 
\lower0.71ex\hbox{{$\scriptstyle \sim$}}}}
\def\gs{\mathrel{\raise0.27ex\hbox{$>$}\kern-0.70em 
\lower0.71ex\hbox{{$\scriptstyle \sim$}}}}
\usepackage{natbib}
\usepackage{epsfig}

\newcommand{\um}{12}
\newcommand{\cit}{1}
\newcommand{\lbl}{2}
\newcommand{\stockholm}{5}
\newcommand{\lpnhe}{6}
\newcommand{\yale}{9}
\newcommand{\lam}{8}
\newcommand{\upenn}{10}
\newcommand{\ucb}{11}
\newcommand{\stsci}{12}
\newcommand{\cppm}{7}
\newcommand{\iu}{13}
\newcommand{\aas}{16}
\newcommand{\uc}{17}
\newcommand{\cambridge}{18}
\newcommand{\cea}{19}
\newcommand{\ipnl}{20}
\newcommand{\slac}{4}
\newcommand{\fnal}{3}
\newcommand{\anu}{15}
\newcommand{\jpl}{14}

\begin{document}
\title{Supernova Acceleration Probe: Studying Dark Energy with Type~Ia
Supernovae\\
{\it A White Paper to the Dark Energy Task Force}}

\renewcommand{\thefootnote}{\fnsymbol{footnote}}
\author{J.~Albert\altaffilmark{\cit},
G.~Aldering\altaffilmark{\lbl},
S.~Allam\altaffilmark{\fnal},
W.~Althouse\altaffilmark{\slac},
R.~Amanullah\altaffilmark{\stockholm}, J.~Annis\altaffilmark{\fnal},
P.~Astier\altaffilmark{\lpnhe},
M.~Aumeunier\altaffilmark{\cppm,\lam},
S.~Bailey\altaffilmark{\lbl},
C.~Baltay\altaffilmark{\yale}
% ...
}

\email{saul@lbl.gov}
\email{melevi@lbl.gov}

\altaffiltext{\cit}{California Institute of Technology}
\altaffiltext{\lbl}{Lawrence Berkeley National Laboratory}
\altaffiltext{\fnal}{Fermi National Accelerator Laboratory}
\altaffiltext{\slac} {Stanford Linear Accelerator Center}
\altaffiltext{\stockholm}{University of Stockholm}
\altaffiltext{\lpnhe}{LPNHE, CNRS-IN2P3, Paris, France}
\altaffiltext{\cppm}{CPPM, CNRS-IN2P3, Marseille, France}
\altaffiltext{\lam}{LAM, CNRS-INSU, Marseille, France}
\altaffiltext{\yale}{Yale University}
\altaffiltext{\upenn}{University of Pennsylvania}
\altaffiltext{\ucb}{University of California at Berkeley}
\altaffiltext{\um}{University of Michigan}
\altaffiltext{\stsci}{Space Telescope Science Institute}
\altaffiltext{\iu}{Indiana University}
\altaffiltext{\jpl}{Jet Propulsion Laboratory}
\altaffiltext{\anu}{The Australian National University}
\altaffiltext{\aas}{American Astronomical Society}
\altaffiltext{\uc}{University of Chicago}
\altaffiltext{\cambridge}{Cambridge University}
\altaffiltext{\cea}{CEA, Saclay, France}
\altaffiltext{\ipnl}{IPNL, CNRS-IN2P3, Villeurbanne, France}

\section{Executive Summary}
The Supernova Acceleration Probe (SNAP) will use Type Ia supernovae
(SNe~Ia) as distance indicators to measure the effect of dark energy
on the expansion history of the Universe.  (SNAP's weak-lensing program
is described in a separate White Paper.)  The experiment exploits
supernova distance measurements up to their fundamental systematic
limit; strict requirements on the monitoring of each supernova's
properties leads to the need for a space-based mission. Results from
pre-SNAP experiments, which characterize fundamental SN~Ia properties,
will be used to optimize the SNAP observing strategy to yield data,
which minimize both systematic and statistical uncertainties.  With
early R\&D funding, we have achieved technological readiness and the
collaboration is poised to begin construction.  Pre-JDEM AO R\&D
support will further reduce technical and cost risk.  Specific details
on the SNAP mission can be found in
\citet{omnibus:2004, Aldering:Widefield}.
\subsection{Overview of Goals and Techniques}
The primary goal of the SNAP supernova program is to provide a dataset
which gives tight constraints on parameters which characterize the
dark-energy, e.g.\ $w_0$ and $w_a$ where $w(a)=w_0+w_a(1-a)$. SNAP
data can also be used to directly test and discriminate among specific
dark-energy models.  We will do so by building the Hubble diagram of
high-redshift supernovae, the same methodology used in the original
discovery of the acceleration of the expansion of the Universe that
established the existence of dark energy
\citep{nature98,Garnavich98,Riess_acc_98, 42SNe_98}.

The SNAP SN~Ia program focuses on minimizing the systematic floor of the
supernova method through the use of char\-ac\-ter\-ized supernovae
that can be sorted into subsets based on subtle signatures of
heterogeneity.  Subsets may be defined based on host-galaxy
morphology, spectral-feature strength and velocity, early-time
behavior, inter alia. Independent cosmological analysis of each subset of
``like'' supernovae can be compared to give confidence that the
results are free from significant systematics.  Conversely, analysis
{\it between\/} supernova subsets at the same redshift  can
identify further systematics controls.  While theories of the
supernova progenitor and explosion mechanism can guide the
establishment of subset criteria, such understanding is not required
-- only comprehensive measurements are -- for robustness of the
cosmological results.  The level of robustness is tied to the quality
of data with which supernovae are distinguished.  Statistical mission
requirements are fundamentally bound by the systematic limitations of
the experiment.

\subsection{Description of the Baseline Proposal}
\begin{figure}
\epsscale{.8}
\plotone{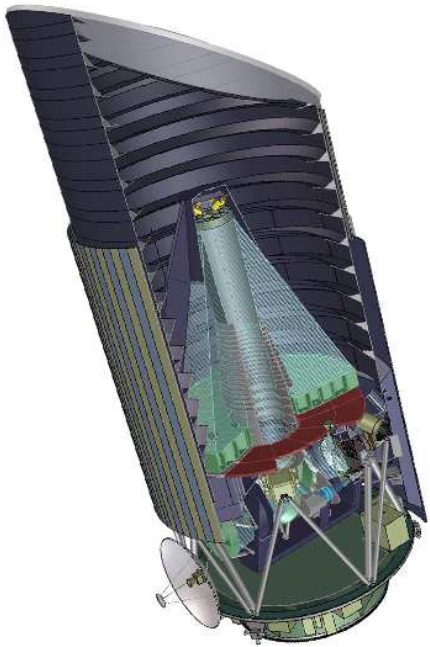}
\epsscale{1}
\caption{Cutaway view of the reference SNAP design.\label{cut:fig}}
\end{figure}

\begin{figure}
\plotone{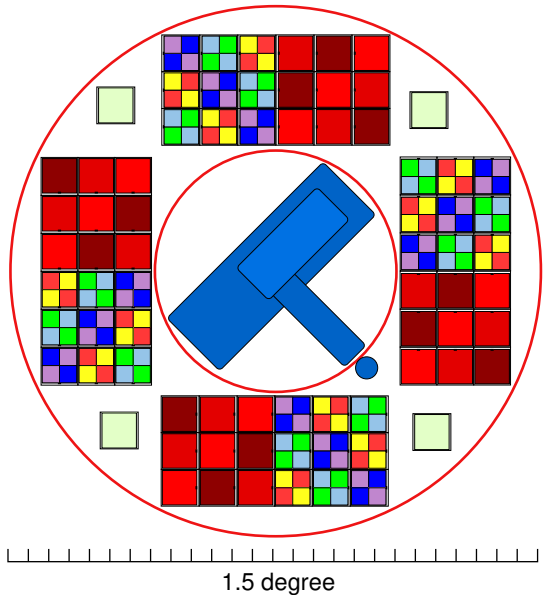}
\caption{The SNAP focal plane working concept. The two-axis symmetry of the imager filters allows any 90$^\circ$ rotation to scan a fixed strip of the sky and measure all objects in all nine filter types. 
The imager covers 0.7 square degrees.  Underlying the filters, there
are 36 2k x 2k HgCdTe NIR devices and 36 3.5k x 3.5k CCDs on a 140K
passively-cooled focal plane.  The six CCD filter types and three NIR filters
are arranged so that vertical or horizontal scans of the array through
an observation field will measure all objects in all filters.  The
false colors indicate filters with the same bandpass.  The central
rectangle and solid circle are the spectrograph body and its light
access port, respectively. The spectrum of a supernova is taken by
placing the star in the spectrograph port by steering the
satellite. The four small, isolated squares are the star guider
CCDs. The inner and outer radii are 129 and 284 mm,
respectively.\label{fplane:fig}}
\end{figure}

\begin{figure}[h]
\plotone{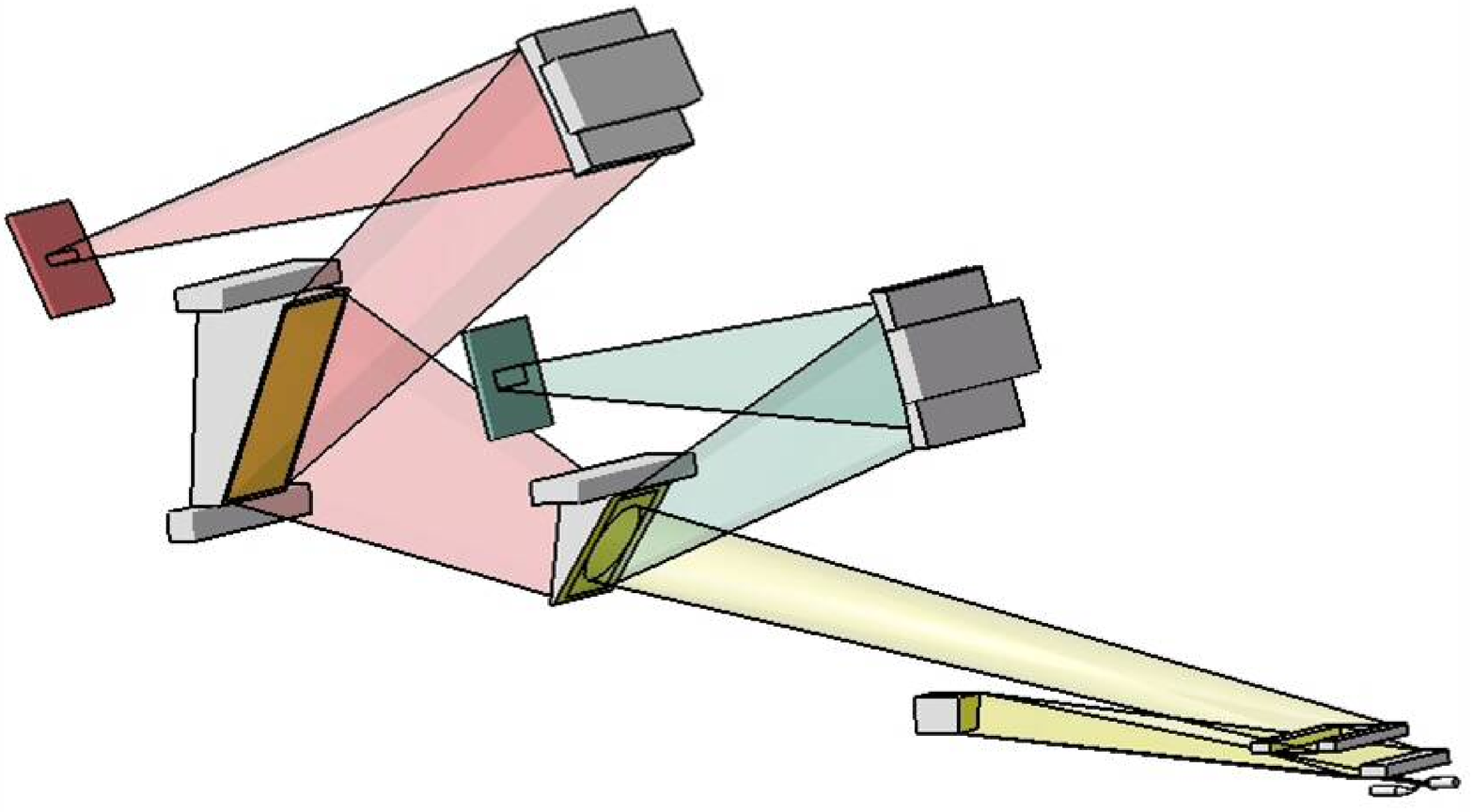}
\caption{
A schematic spectrograph optical design. The beam is going out from
the slicer (on the bottom right) to a prism disperser back faced
coated by a dichroic. The visible light (blue path) is reflected and
the IR light (in red) continue to a second prism used to reach the
required spectral dispersion. The two beams are therefore focused
on two detectors.  The dimensions of the spectrograph are approximately
$400 \times 80 \times 100$ mm.  See \citet{Ealet02} for more details.
\label{sblock:fig}}
\end{figure}

SNAP is a 2-m space telescope (shown in Figure~\ref{cut:fig})
with a 0.7 square-degree instrumented
focal plane imager (shown in Figure~\ref{fplane:fig}) and a
low resolution $R \sim 100$
integral-field-unit spectrograph (schematic shown in
Figure~\ref{sblock:fig}).  Both instruments are sensitive to
the 0.4 -- 1.7 $\mu$m wavelength range.  The two SNAP deep-survey
fields each subtends 7.5 square degrees and each is observed 
approximately every four
days over the time-scales of supernova light-curve evolution in each
of nine passbands.  The passbands have resolution $\sim$4.5 and are
logarithmically distributed over the full wavelength range.  Targeted
spectrographic observations are made of discovered supernovae with
exposure times tuned to the source flux.


The magnitude depths for individual scans and co-added images of the
SNAP supernova fields are calculated for each filter.
Table~\ref{depth:tab} shows the limiting $S/N=5$ AB magnitude for each
filter in the ``Deep'' supernova survey, for point sources that are
not contaminated with cosmic rays. The limiting magnitude for any
given point is probabilistic, due to the random occurrence of cosmic
rays.
In a single scan of the deep survey, 77\%
of point sources will have no cosmic-ray contamination in any of the four
dithered images while 98\% will have one or fewer cosmic-contaminated
image, only slightly reducing signal-to-noise.

\begin{deluxetable}{cccccccc}
\tablewidth{0pt}
\tablecaption{The SNAP
AB magnitude survey depth for a point source $S/N=5$.\tablenotemark{a}
\label{depth:tab}}
\tablehead{
Filter & $\lambda_{eff}$(\AA) & $\Delta \lambda$(\AA)& \multicolumn{2}{c}{Deep Survey (AB mag)} & Wide-field Survey & Panoramic\\ \cline{4-5}
& & &Per Scan & Co-added Scans& (AB mag)&(ABmag)}
\startdata
      1&        4400&        1000&   27.9&   30.6&   28.3&   26.7\\
      2&        5060&        1150&   27.9&   30.6&   28.3&   26.8\\
      3&        5819&        1322&   27.8&   30.5&   28.2&   26.8\\
      4&        6691&        1520&   27.8&   30.4&   28.1&   26.8\\
      5&        7695&        1749&   27.8&   30.4&   28.1&   26.8\\
      6&        8849&        2011&   27.7&   30.3&   28.0&   26.7\\
      7&       10177&        2313&   27.5&   30.1&   27.8&   26.6\\
      8&       11704&        2660&   27.5&   30.1&   27.8&   26.6\\
      9&       13459&        3059&   27.4&   30.0&   27.7&   26.5\\ 
\enddata
\tablenotetext{a}{Random cosmic-ray hits make the $S/N$ for a given position probabilistic; see text.
The choice of filter set is currently subject to optimization studies;
the filters and depths presented here are meant to be illustrative.}
\end{deluxetable}

\section{Precursor Observations}
Any precursor mission that characterizes the fundamental properties of
SNe~Ia and their applicability as distance indicators is important for
optimizing the observing strategy and data analysis of SNAP.

Low-redshift surveys provide high signal-to-noise, comprehensive
observations of well-char\-ac\-ter\-ized SNe~Ia and could provide
specific requirements for the observation of light-curve or
spectroscopic features linked to SN~Ia heterogeneity and homogeneity.
Heterogeneous features provide a handle to probe the systematic
limitations of our distance indicator and form the criteria by which
``like'' supernova subsets are defined.  Conversely, current distance
measurements to supernovae may be found not to be optimal; a different
mix of observations may accentuate supernova homogeneity and increase
the probative power of a single event.  Finally, these surveys will
provide an expanded library of spectra for K-corrections.

Ongoing high-redshift supernova searches and SNAP probe similar depths
and employ similar observing strategies.  Precursor missions establish
high-redshift supernova rates and fluxes which will impact the SNAP survey
strategy.  The search and follow-up of ``rolling'' supernova searches
(continuous repeated observing of the target fields)
provide a testbed for SNAP discovery and trigger for spectroscopic
follow-up.  These searches may indicate that specific observables are
linked to redshift-dependent evolution in supernova populations.
Finally, these searches can access and allow the study of the
supernova-frame UV which has unexploited potential for aiding in
supernova cosmology measurements.

\section{Error budget}
SNe~Ia have already proved to be excellent distance indicators
for probing the dynamics of the Universe. In order to move toward the
era of precision cosmology, a new data set must both address potential
systematic uncertainties and provide sufficient statistical power to
accurately determine dark-energy properties, especially the time
variation of the equation of state.
\subsection{Sources of Systematic Uncertainty}
\label{exec_uncertain}


High-redshift supernova searches have been proceeding since the late
1980's \citep{Norgaard:1989,Couch_etal:1989,
Perlmutter_etal:1995,7results97,nature98,
42SNe_98,Schmidt_98,Riess_acc_98,Tonry_etal:2003,Knop_etal:2003,
Riess_and_others:2004}.  Particularly since the discovery of the
accelerating expansion of the Universe, the high-redshift supernova
methodology for measuring cosmological parameters has been critically
scrutinized for sources of systematic uncertainty.  Below
(and summarized in Table~\ref{sys:tab}) is a lists of
these sources where we highlight in {\bf bold} the features specific
to SNAP that allow systematic control.  Generically, the need for
high signal-to-noise observations over a broad wavelength range
for $0.1<z<1.7$ supernovae, as
well as the truncation of ground-observatory light curves of high-redshift
supernova at high-Galactic latitude, point to the necessity of a space
mission.

\begin{deluxetable}{lp{4.5in}}
\tablecaption{SNAP Control of Sources of Astrophysical 
Systematic Uncertainty
\label{sys:tab}}
\tablehead{Systematic & Control}
\startdata
Supernova evolution & Use of only the best-characterized supernovae
subclassified through high signal-to-noise light curves and
spectrum(a)\\ K Correction & Tuning of the SNAP passbands and
accumulation of a library of supernova spectra\\ Galactic extinction &
Search fields in the direction of the low-extinction Galactic caps\\
Cross-passband calibration & A SNAP program to construct a set of
high-precision flux standards\\ Host-galaxy dust &
Wavelength-dependent absorption identified with high signal-to-noise
multi-band photometry spanning a broad wavelength range\\
Gravitational lensing & Measure the average flux for a large number of
supernovae in each redshift bin\\ Non-Type~Ia contamination &
Classification of each event through the 6150\AA\ SiII and S spectral
features\\ Malmquist bias & Supernovae discovered early with high
signal-to-noise photometry\\ Gray dust & Photometry and spectroscopy
for supernovae at wavelengths where gray dust is no
longer achromatic to search for gray dust at moderate redshift
\enddata
\end{deluxetable}

\subsubsection{Evolution of the Supernova Sample as a Function of Redshift}
The luminosity function of supernovae may evolve as a function of
redshift resulting in corresponding biases in distance-modulus
determinations \citep{Perlmutter_etal:1995}.  Although
like supernovae have the same redshift-independent luminosity,
their relative rate of occurrence may evolve in redshift.

SNAP addresses this problem by identifying subsets of supernovae
``identical'' in luminosity and color, applying the dark-energy
analysis to each, and optimally combining the subset results into
final cosmology measurements.  We exploit the fact that supernovae
cannot change their brightness in one measured wavelength range
without affecting brightness somewhere else in the spectral time
series --- an effect that is well-captured by expanding atmosphere
computer models \citep[e.g.][]{hofkhoklc96, Lentz00, Linder:Hoeflich}.
Well-characterized supernovae can be sorted into subsets based on
subtle signatures of heterogeneity.  Subsets may be defined based on
host-galaxy morphology, spectral-feature strength and velocity,
early-time behavior, and light-curve characteristics.

There are a number of supernova parameters that are observed to
exhibit heterogeneity beyond the single-parameter description.  The
following
features need to be monitored to check for systematic bias
in supernova distances.

{\it Premaximum spectral screening:} 
Broad-light-curve supernovae can be distinguished by having either
peculiar (e.g.\
SN~1991T) or ``normal'' (e.g.\ SNe 1999ee and 2000E) pre-maximum spectra.  In
addition, there are extremely peculiar singleton events technically
classified as SNe~Ia which bear little observational resemblance to
their more standard counterparts.  These include SN 2000cx
\citep{Li:SN2000cx, Candia:2003}, SN 2001ay \citep{Phillips:2003}, SN
2002cx \citep{Li:SN2002cx}, and SN 2002ic \citep{Deng:2002ic}.  These
extreme events are easily identified through their premaximum peculiar
spectra, Hydrogen emission, and high expansion velocities.

{\it Spectral-feature velocities:} The diversity in the velocity of
the 6100\AA\ SiII feature at fixed epochs has been noted
\citep{bvdb93,Hatano_etal:2000}.  \citet{Garavini:2004} and
\citet{Benetti:2004} have
examined the velocity as a function of supernova phase and find that
large subset of objects tightly cluster within a $10^3$ km s$^{-1}$
range for any given epoch.  A smaller subset with much
higher velocity dispersion  can be partially distinguished through
accurate measurements of the SiII feature velocity.

{\it Spectral-feature velocity evolution:}
\citet{Benetti:2004} find that supernovae cluster when parameterized by
the rate of change of the SiII feature velocity $\dot{v}$ and
luminosity.  Resolution of the velocity gradients requires measurement
of $\dot{v}$ to a few km s$^{-1}$ d$^{-1}$ from spectra between peak
and 40 SN-frame days after peak brightness.

{\it Spectral-line-ratio:} At $t<-10$, the ratio of
line 
depths of SiII($\lambda$ 6150\AA) and SiII($\lambda$ 5800\AA), R(SiII)
\citep{Nugent95}, has a range greater than 0.1.  The subpopulation of
supernovae with high velocity gradients also have distinct R(SiII)
evolution; they sharply decrease until $t=-5$ and flatten out whereas
other supernovae show no evolution for the same time range
\citep{Benetti:2004}.

{\it Light-curve evolution:}
There are a series of supernovae that exhibit systematic deviations in
light-curve shape and color
\citep[e.g.][]{Hamuy:1996,Richmond:sn94d, Krisciunas:2003,Pignata:SN2002er}.  The ranges displayed in the optical
light curves can be summarized as follows: 1 mag 10 days before
maximum and 0.2 mag 5 days before maximum, 0.1 mag 20-30 days after
maximum, and 0.2-0.3 mag 60 days after maximum.

{\it Color:}
There is evidence that color is a
second parameter correlated with peak magnitude
\citep{Tripp98, Saha:1999, Tripp:1999, Parodi:2000}.
Dust extinction and intrinsic supernova
color can be distinguished as supernovae get brighter in $I$ as
$V-I$ color becomes red, the opposite effect from dust.
Color measurements whose propagated uncertainties in the
extinction correction are less than the intrinsic $B$ peak magnitude
dispersion correspond to $B-V$ measured to 0.07 mag and $V-I$ to 0.5 mag.

{\it Polarization:} Normal- and broad-light-curve supernovae have
demonstrated varying degrees of polarization \citep{Howell:2001}.
Detection of polarization is extremely difficult for all but the
nearest supernovae and is unrealistic to require in any cosmological
survey.  Spectroscopy may then serve as a proxy for polarization
measurements through their correlation with high and quickly evolving
SiII velocities.

{\bf The SNAP instrumentation suite is designed to provide the
data necessary for the subclassification process.  SNAP will obtain
broadband photometry over the full supernova-frame optical region,
with fine light-curve sampling tuned to natural supernova time scales.
The SNAP spectrograph also covers the supernova-frame optical region
with resolution tuned to the widths of the supernova
features.}

We expect that any residual luminosity and color biases will be
smaller than the dispersion observed in the local supernova sample
since this represents heterogeneous and
diverse supernova-progenitor environments \citep{Branch:2001}.  This
assumption will be tested by intensive precursor
supernova-characterization experiments and SNAP itself.

\subsubsection{K-correction}
A given observing passband is sensitive
to different supernova-rest-frame spectral regions.  K-corrections are
used to put these differing photometry measurements onto a consistent
rest-frame passband \citep*{Kim_kcorr96,Nugent:2002}. {\bf The SNAP
filter-set number, density, and shapes are tuned} to give $<0.02$ mag
dispersion for the full catalog of currently available SN~Ia spectra
\citep{Davis:2005}.  The situation will improve further for the
confined analysis of supernova subsamples with the expanded spectral
library provided by ongoing surveys and SNAP itself.

\subsubsection{Galactic Extinction}
Extinction maps of our own Galaxy are uncertain by $\sim$ 1--10\%
depending on direction \citep*{Schlegel_dust_98}.  Supernova fields
can be chosen toward the low-extinction Galactic caps where
uncertainty can then be controlled to $<0.5$\% in brightness. These
levels of uncertainty have negligible effect on the measurement of
$w_0$ and $w_a$ \citep{Kim&Miquel:2005}.  Spitzer observations will
allow an improved mapping between color excesses (e.g.\ of Galactic
halo subdwarfs in the SNAP field) and Galactic extinction by dust
\citep{Lockman:2005}.

\subsubsection{Cross-Passband Calibration}
Uncertainties in calibration are one of the important limiting factors
in probing the dark energy with supernovae.  Distance measurements
fundamentally rely on the comparison of fluxes in different passbands:
for determining the relative distances of supernovae at different
redshifts and for color measurements for determining the dust
absorption of a single supernova.

The relationship between passband calibration and $w_0$ and $w_a$ is
strongly dependent on the specifics of the cross-band correlations.
Simulation studies \citep{Kim&Miquel:2005} demonstrate the
strict fundamental limits in the measurement of dark-energy
parameters due to uncorrelated cross-band calibration uncertainty.
Significantly larger correlated
uncertainties can be tolerated \citep{Kim_etal:03}.
These studies provide requirements for {\bf the SNAP calibration program.}

\subsubsection{Extinction by Host-Galaxy Dust}
Extinction from host-galaxy dust can significantly reduce the observed
brightness of a discovered supernova.  {\bf SNAP's 9-band photometry
will identify the properties of the obscuring dust and gas, and thus the
amount of extinction suffered by individual supernovae.} Incoherent
scatter in dust properties yields no bias but increased statistical uncertainty
in the dark-energy parameters \citep{Miquel_Linder:2004}.  Systematic
host-galaxy extinction uncertainty can be introduced if an incorrect
mean dust-extinction law is used in the analysis.  This can be
minimized by {\bf distinguishing and
excluding highly-extincted supernovae from the
analysis}.  To give a quantitative example, suppose that the
\citet{card89} model accurately describes host-galaxy extinction 
and an $R_V=3.1$ dust is used in the analysis and the $4400$ --
$5900$\AA\ rest-frame color is used to determine $A_V$.  We calculate
that if drift in the mean of the true dust extinction models is
constrained to have $2.4<R_V<4.4$, the vast majority of supernova
which have $A_B<0.1$ \citep*{Hatano98} would have distance-modulus
bias of  $<0.02$ mag.

\subsubsection{Gravitational Lensing} 
Inhomogeneities along the supernova line of sight can gravitationally
magnify or demagnify the supernova flux and shift the mode of the
supernova magnitude distribution by $\ls 3$\% depending on redshift
\citep{Holz&Linder:2004}.  Gravitational lensing magnification is
achromatic and unbiased in flux so effectively contributes a
statistical uncertainty that can be reduced below 1\% in distance 
if {\bf large numbers ($\gs 70$) of supernovae are obtained per 0.1
redshift bin}.  The effective flux dispersion induced by lensing goes as
$\sigma=0.088z$ \citep{Holz&Linder:2004}.

\subsubsection{Non-SN Ia Contamination}
Other supernova types are on average
fainter than SNe Ia and their contamination could bias their Hubble
diagram \citep{42SNe_98}.  Observed supernovae must be positively
identified as SN~Ia.  As some Type~Ib and Ic supernovae have spectra
and brightnesses that otherwise mimic those of SNe~Ia
\citep{Filippenko:1997}, {\bf a spectrum
covering the defining rest frame Si~II 6150\AA\ feature and SII
features for every supernova at maximum} will provide a pure sample.

\subsubsection{Malmquist Bias}
A flux-limited sample preferentially detects the intrinsically
brighter members of any population of sources.  The amount of
magnitude bias depends on details of the search but can reach the
level of the intrinsic magnitude dispersion of a standard candle
\citep{Teerikorpi:1997}.  Directly correcting this bias would rely on
knowledge of the SN Ia luminosity function, which may change with
lookback time.  {\bf A detection threshold fainter than peak by at
least five times the intrinsic SN~Ia luminosity dispersion} ensures
sample completeness with respect to intrinsic supernova brightness,
eliminating this bias
\citep{Kim_etal:03}.


\subsubsection{Extinction by Hypothetical Gray Dust}
A systematics-limited experiment must account for speculative but
reasonable sources of uncertainty such as gray dust.
As
opposed to normal dust, gray dust is postulated to produce
wavelength-independent absorption in optical bands
\citep{Aguirre_99,Aguirre&Haiman:2000}.  Although models for gray-dust
grains dim blue and red optical light equally, near-infrared
light ($\sim$1-2 $\mu$m) is less affected.  {\bf Cross-wavelength
calibrated spectra extending to wavelength regions where ``gray'' dust
is no longer gray} will detect and correct for the hypothetical large-grain
dust.  Numerous observations already severely limit the cosmological
effects of such a component  \citep*{Goobar:2002}.

\subsection{Statistical Supernova Sample}
\label{syserrmod:sec}
The statistical supernova sample appropriate for experiments with the
presence of systematic uncertainty has been considered in the
Fisher-matrix analysis of \citet{hut:2003} and
validated with our independent Monte Carlo analysis. They examined
the measurement precision of dark-energy parameters as a function of
redshift depth considering 2000 SNe~Ia measured in the range $0.1\le
z\le z_{\rm max}$, along with 300 low-redshift SNe~Ia from the Nearby
Supernova Factory \citep{Aldering02}.  
They concluded that a SN~Ia sample extending to redshifts of $z \sim
1.7$ is crucial for realistic experiments in which some systematic
uncertainties remain after all statistical corrections are applied.
Ignoring systematic uncertainties can lead to claims that are too
optimistic.
 
Statistically, $\sim$2000
SNe~Ia well-characterized ``Branch-normal'' supernovae provide
systematic-uncertainty limited measurements of the dark-energy
parameter $w_a$.  These represent the cream, $z<1.7$ supernovae
of the
$\sim 10000$ total and $\sim 4000$ well-characterized SNAP SNe~Ia, that
can provide robust cosmological measurements (e.g.\ non-peculiar,
low-extinction).  This large sample is necessary to allow model-independent
checks for any residual systematics or refined standardization
parameters, since the sample will be subdivided in a
multidimensional parameter space of redshift, light curve-width, host
properties, etc.  The large number of lower signal-to-noise light
curves from supernovae at $z>1.7$ will provide an extended redshift
bin in which to check for consistency with the core sample.

We require a
statistical measurement uncertainty for the peak magnitude corrected
for extinction and shape inhomogeneity roughly equal to the current
intrinsic magnitude dispersion of supernovae, $\sim 0.15$ mag.

\section{Quantification of Dark Energy}
\begin{figure}[h!]
\plotone{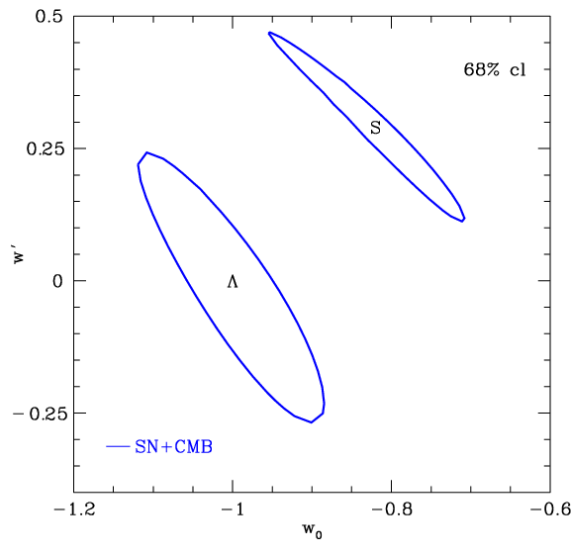}
\caption{The 68\% confidence region in the dark-energy parameters
for fiducial cosmological constant (labeled with $\Lambda$) and
supergravity (labeled with S) universes.  The blue solid curves show
the results of the supernova simulation and associated
systematic-error model, with a Planck prior on the distance
to the surface of last scattering, and an assumed flat
universe. We define $w'\equiv
-dw/d\ln a|_{z=1}$. See \citet{snap:detf:wl}
 for constraints including the weak lensing half of the mission. 
\label{w0w':fig}}
\end{figure}

Expected measurement precisions of the dark-energy parameters $w_0$
and $w'(=w_a/2$) for the SNAP supernova program are summarized in
Table~\ref{sci_errors.tab} and shown in Figure~\ref{w0w':fig}.  Since
SNAP is a systematics-limited experiment, any improvements in our
projection of systematic uncertainty will be
directly reflected in improvements in our $w_0$--$w'$ measurements.
We determine precisions for two fiducial flat universes with
$\Omega_M=0.3$, one in which the dark energy is attributed to a
cosmological constant and the other to a SUGRA-inspired dark-energy
model \citep{Brax:1999}.

\begin{deluxetable}{cccccc}
\tablewidth{0pt}
\tablecolumns{5}
\tablecaption{SNAP 1-$\sigma$ uncertainties in dark-energy parameters, with conservative systematics for the supernova and the shear power-spectrum measurement from
a 1000 sq.\ deg.\ one-year weak-lensing survey.  Note: These uncertainties are systematics-limited, not statistics limited.\label{sci_errors.tab}}
\tablehead{
& $\sigma_{\Omega_M}$ & $\sigma_{\Omega_w}$ & $\sigma_{w_0}$ &
$\sigma_{w'}$

}
\startdata
Fiducial Universe: flat, $\Omega_M=0.3$, Cosmological Constant dark energy \\
\cline{1-1}\\
SNAP SN; $w=-1$& 0.02 &  0.05 & \nodata & \nodata  \\
SNAP SN + WL; $w=-1$
                       & 0.01  &  0.01   & \nodata &\nodata\\
SNAP SN;$\sigma_{\Omega_M}=0.03$ prior; flat; $w(z)=w_0+2w'(1-a)$
                       & \nodata &  $0.03$ & $0.09$ & $0.31$  \\
SNAP SN;Planck prior; flat; $w(z)=w_0+2w'(1-a)$
                       & \nodata  &  $0.01$   &$0.09$ & $0.19$  \\
SNAP SN + WL; flat; $w(z)=w_0+2w'(1-a)$
                       & \nodata  &  $0.005$   & $0.05$&$0.11$\\
\\
Fiducial Universe: flat, $\Omega_M=0.3$, SUGRA-inspired dark energy\\
\cline{1-1}\\
SNAP SN; $\sigma_{\Omega_M}=0.03$ prior; flat; $w(z)=w_0+2w'(1-a)$
                       & \nodata  &  $0.03$   & $0.08$&$0.17$\\
SNAP SN; Planck prior; flat; $w(z)=w_0+2w'(1-a)$
                       & \nodata  &  $0.02$   & $0.09$&$0.13$\\
SNAP SN + WL; flat; $w(z)=w_0+2w'(1-a)$
                       & \nodata  &  $0.005$  & $0.03$&$0.06$\\
\enddata
\tablecomments{Cosmological and dark-energy parameter
precisions for two fiducial
flat universes with $\Omega_M=0.3$, one in which the dark energy is
attributed to a Cosmological Constant and the other to a
SUGRA-inspired dark-energy model.  The parameter precisions then
depend on the choice of data set, priors from other experiments
(e.g.\ Planck measurement of the distance to the surface of last
scattering),
assumptions on the flatness of the universe,
and the model for the behavior of $w$.
SNAP is a systematics-limited experiment so any improvements in our
projection of systematic uncertainty will be
directly reflected in improvements in our $w_0$--$w'$ measurements.
In this paper we define $w'\equiv -dw/d\ln a|_{z=1}$.
SNAP can accumulate a wide-field weak-lensing survey at a rate of
1000 sq.\ deg.\ per year of observation.}
\end{deluxetable}

If the universe is taken to be spatially flat, the SN data (with
Planck prior on the distance to the surface of last scattering)
determine $\Omega_M$ to 0.01.  The present value of the
dark energy equation of state is constrained within 0.09 and the
physically crucial dynamical clue of the equation of state time
variation is bounded to within 0.19.
Note that these constraints are {\it systematics limited};
much tighter constraints would be obtained if systematics were
ignored.  Table~\ref{sci_errors.tab} also highlights the significant
value added
by measurements of the shear power spectrum with the
complementary SNAP weak-lensing program.

The precise, homogeneous, deep supernova data set can alternatively
be analyzed in a non-parametric, uncorrelated bin, or
eigenmode method \citep{Huterer&Starkman:2003,Huterer&Cooray:2004}.  One
can also obtain the expansion history $a(t)$ itself, e.g.\
\citet{Linder:2003}.

\section{Project Risks and Strengths}
The proposed experiment is a self-contained supernova program which is
desensitized to cross-experimental calibration, one of the largest
sources of systematic uncertainty.  The SNAP instrumentation suite
provides access to any supernova observable anticipated to be
important for reducing uncertainties in supernova distances.

The SNAP collaboration is large and diverse, which
has allowed for substantial efforts to be supported in science reach,
computer framework and simulation, calibration approaches, and
photo-detector technology development and testing.
Our ongoing R\&D efforts have placed us at a point of technological
readiness to commence construction of the experiment.

The major risk to the SNAP collaboration is that a stretch in the JDEM
decision making process and follow-on project funding will lead
to a dissipation of the present efforts as people drift off to
other intermediate projects.

\section{Requisite technology R\&D}
A few areas of the SNAP mission implementation require development to
achieve flight readiness. No new technologies are required, rather
particular instances of the technology as chosen by SNAP require some
maturation. At present, the SNAP R\&D program is primarily funded by
the Department of Energy (DOE), with additional support to
collaborators from NSF, NASA, CNES, and CNRS/IN2P3 for related
activities.  The SNAP mission architecture has undergone numerous
studies (GSFC/IMDC, GSFC/ISAL, and JPL/Team-X), has been reviewed
several times by external teams of experts at the request of DOE and
NSF, and has been extensively presented and discussed at a variety of
international conferences.  Thus, SNAP is a relatively well-advanced
concept, and the internally self-consistent reference model presented
here is now ready for more refined optimization.

R\&D has focused on 1) SNAP-specific detector development, both visible
and near infrared, 2) integral field unit spectrograph design, 3) cold
front-end electronics, 4) calibration procedures and supporting flight
hardware, 5) achievable telescope point stability studies, 6) on-board
data handling, and 7) achievable telemetry rate studies. All of these
areas are maturing at a rapid rate and there appear to be no show
stoppers for a SNAP JDEM response.

Visible detectors: The visible portion of the imager is an array of 36
silicon CCDs with extended red response and improved radiation
tolerance. SNAP will materially benefit from an advanced technology
based on fully depleted backside illuminated high resistivity
silicon.  The development work is DOE funded and is taking
place at LBNL and at commercial foundries.  We are also developing
custom integrated circuits for the electronic readout.

NIR detectors: The NIR portion of the imager requires 36 2kx2k format
devices. We are evaluating HgCdTe devices available from two
vendors.  Development work funded by DOE is under way to improve the
dark current, read noise, and quantum efficiency.  We are also
exploring InGaAs as a backup photoactive material. One NIR vendor will
provide a cold front-end ASIC for their detectors. For the other, we
are leveraging the work we have done for the CCDs to provide cold
front-end ASICs.

Spectrograph: This is a simple two-prism design with no technology
challenges except its integral field unit (IFU).  An image-slicer
IFU with the
requisite size and efficiency have been built for ground-based
astronomy, and will be flown as part of the JWST NIRSPEC.  The image
slicer R\&D is funded by CNES and CNRS/IN2P3.

Telescope: This has been studied by the Goddard/ISAL and four
prospective space optics suppliers.  Two industry study contracts are
in place to evaluate the design, manufacturability, and testability of
the telescope. Studies and vendor discussions have shown that the
pointing stability requirements can be met with today's spacecraft
technology.

Data handling: We have developed an on-board detector data handling
concept where each detector operation, data
processing, and storage is done in parallel. This allows for
graceful degradation of the focal plane performance at the detector
level. The
large integrated data storage required and the way it is used
favors FLASH RAM technology. Initial radiation testing indicates
that this is a viable approach.

Telemetry: We require a 150 Mbit/s Ka-band downlink.  While the ground
portion of this system can be bought ``off the shelf'', and heritage
Ka-band TWTA transmitters of the required power are readily available,
the 150 MHz modulator is a development item. Signal processing
upgrades will be required to existing ground stations to handle 150
Mbit/s.

Calibration: An important feature of the supernova-cosmology analysis
is its dependence on the {\it relative} brightness of Type Ia
supernovae.  The SNAP calibration program must thus precisely control
the absolute {\it color} calibration in the $0.4 - 1.7$ $\mu$m
observing window.  We plan to establish a set of astronomical sources
with known physical fluxes as ``fundamental standards''.  A network of
fainter primary and secondary standard stars, accessible to SNAP and
larger ground-based telescopes, are to be established with respect to
the fundamental standards.  The requirements on the accuracy of the
calibration standards are defined by their propagated effect on
$w_0$ and $w_a$.

\section{Facility Access}
To receive the SNAP science data, access
to the NASA Deep Space Network or other NASA ground
stations is desired. Upgrades to process 150 Mbit/s demodulation are
required.

\section{Timeline}
With the considerable detector and mission development which has
already been completed, the SNAP collaboration is ready to begin
construction of the experiment.


\end{document}